\documentclass{jfm}
\usepackage{amsmath,appendix}
\usepackage{amssymb}
\usepackage{graphicx,color}
\usepackage{natbib}
\usepackage{epstopdf, epsfig}
\usepackage{euscript}

\newcommand{\beq}{\begin{equation}}
\newcommand{\eeq}{\end{equation}}

\newcommand{\bfb}{\mbox{\boldmath $b$}}

\newcommand{\bfk}{\mbox{\boldmath $k$}}

\newcommand{\bfv}{\mbox{\boldmath $v$}}
\newcommand{\bfx}{\mbox{\boldmath $x$}}

\newcommand{\bfr}{\mbox{\boldmath $r$}}

\newcommand{\bfB}{\mbox{\boldmath $B$}}
\newcommand{\bfC}{\mbox{\boldmath $C$}}

\newcommand{\bfM}{\mbox{\boldmath $M$}}

\newcommand{\bfU}{\mbox{\boldmath $U$}}
\newcommand{\bfV}{\mbox{\boldmath $V$}}

\newcommand{\bfP}{\mbox{\boldmath $P$}}
\newcommand{\bfQ}{\mbox{\boldmath $Q$}}
\newcommand{\bfxi}{\mbox{\boldmath $\xi$}}
\newcommand{\ex}{\mbox{{\boldmath $e$}}_{1}}
\newcommand{\ey}{\mbox{{\boldmath $e$}}_{2}}
\newcommand{\ez}{\mbox{{\boldmath $e$}}_{3}}

\newcommand{\bfemf}{\mbox{\boldmath ${\cal E}$}}
\newcommand{\cross}{\mbox{\boldmath $\times$}}
\newcommand{\cendot}{\mbox{\boldmath $\cdot\,$}}

\shorttitle{Moffatt drift dynamo}
\shortauthor{N. K. Singh}

\title{Moffatt drift driven large scale dynamo due to $\alpha$
fluctuations with nonzero correlation times}

\author{Nishant K. Singh\aff{1}
  \corresp{\email{nishant@nordita.org}}}

\affiliation{\aff{1}Nordita, KTH Royal Institute of Technology
and Stockholm University, Roslagstullsbacken 23, SE-10691 Stockholm,
Sweden}

\begin{document}

\pagerange{\pageref{firstpage}--\pageref{lastpage}} \pubyear{2013}

\maketitle

\label{firstpage}

\begin{abstract}
We present a theory of large--scale dynamo action in a turbulent flow that has
stochastic, zero--mean fluctuations of the $\alpha$ parameter.
Particularly interesting is the possibility of the growth of the mean
magnetic field due to Moffatt drift, which is expected to be finite
in a statistically anisotropic turbulence.
We extend the
Kraichnan--Moffatt model to explore effects of finite memory of $\alpha$
fluctuations, in a spirit similar to that of \cite{SS14}, hereafter SS14.
Using the first--order smoothing approximation,
we derive a linear integro--differential equation governing the dynamics
of the large--scale magnetic field, which is non--perturbative in the
$\alpha$--correlation time $\tau_{\alpha}$. 
We recover earlier results in the exactly solvable white--noise (WN) limit
where the Moffatt drift does not contribute to the dynamo growth/decay.
To study finite memory effects, we reduce the
integro--differential equation to a partial differential
equation by assuming that the $\tau_{\alpha}$ be small but nonzero and
the large--scale magnetic field is slowly varying. We derive the dispersion
relation and provide explicit expression for the growth rate as a function
of four independent parameters. When $\tau_{\alpha}\neq 0$, we find that:
(i) in the absence of the Moffatt drift, but with finite Kraichnan diffusivity,
only strong $\alpha$--fluctuations can enable a mean--field dynamo
(this is qualitatively similar to the WN case);
(ii) in the general case when also the Moffatt
drift is nonzero, both, weak or strong $\alpha$ fluctuations, can lead to a
large--scale dynamo; and (iii) there always
exists a wavenumber ($k$) cutoff at some large $k$ beyond which the growth rate
turns negative, irrespective of weak or strong $\alpha$ fluctuations.
Thus we show that a finite Moffatt drift can always facilitate large--scale dynamo
action if sufficiently strong, even in case of weak $\alpha$ fluctuations,
and the maximum growth occurs at intermediate wavenumbers.

\end{abstract}

\begin{keywords}
dynamo theory --- magnetohydrodynamics --- turbulence theory
\end{keywords}

\section{Introduction}

The magnetic fields observed in various astrophysical bodies, such as the planets,
the Sun, stars, galaxies etc, are believed to be self--sustained by turbulent dynamos
\citep{Mof78,Par79,KR80,RSS88,Kul04,BS05,Jon11}. In an electrically conducting plasma,
conversion of the kinetic energy into the magnetic energy, without any electric
current at infinity, is known as the dynamo
action, which leads to amplification of weak seed magnetic field.
Magnetic fields exhibit coherence over a range of scales, from smaller to much larger
than the outer scale of the turbulence. Systems lacking ordered motion, such as
the clusters of galaxies, predominantly host a fluctuation dynamo
\citep{Mur04,VE05,KE11,BS13}, whereas those with large--scale motion, such as the Sun,
galaxies etc, also support a large--scale dynamo
\citep[and references therein]{Cha10,Bec12,Cha13a,Cha13b,Cha14}.
Dynamo origin of the galactic magnetic field seems unchallenged.
Helical turbulence has been considered to be the key driver for large--scale dynamo,
which could operate even in systems without mean motion. For example, both, analytical
calculations and numerical simulations, reveal that even in the absence of differential
rotation or mean motion, large--scale magnetic field grows due to helically forced
turbulence by what is known as an $\alpha^2$--effect \citep{BS05}. However, it is not clear
whether astrophysical turbulence has a mean helicity that is large enough to sustain
such a large--scale turbulent dynamo. This brings us to a following natural question:
could the large--scale magnetic fields grow if the helicity of the turbulence vanishes
on average? \cite{Kra76} was the first to study this problem by considering $\alpha$
(which is a measure of mean kinetic helicity of the turbulence) as a stochastic variable,
with zero mean, and demonstrated that the $\alpha$ fluctuations lead to a decrement of
the turbulent diffusivity, and if sufficiently strong, they could give rise to the growth
of mean magnetic field by the process of \emph{negative diffusion}. \cite{Mof78}
generalized this model to include a statistical correlation between the fluctuating
$\alpha$ and its spatial gradient, and found that this contributes a constant drift
velocity to the dynamo action, but it does not affect the dynamo condition.
Both \cite{Kra76} and \cite{Mof78} have essentially considered white--noise $\alpha$
fluctuations, as has been explicitly shown by SS14 who extended previous studies
by also studying the memory effects, giving rise to new interesting mechanisms.

The SS14 model is
limited to fairly low wavenumbers because their FOSA calculation of the mean
EMF ignored turbulent resistivity. The present paper remedies this, obtaining
a new expression for the mean EMF which predicts a significant modification
of dynamo action at intermediate and large wavenumbers.
We show that the inclusion of the resistive term in the present work
to determine mean EMF is a nontrivial extension of SS14 model, leading
to even qualitatively new predictions for the growth rates at
intermediate and large wavenumbers.
Considering deterministic Roberts flows, \cite{Rhe14} discussed the role of
finite memory for the existence of the dynamo.
Based on numerical experiments of passive scalar diffusion and kinematic
dynamos, \cite{HB09} investigate turbulent transport where
the turbulence possesses memory.
These works highlight the importance of the memory effects on the
turbulent transport processes and demonstrate that the turbulent
transport coefficients can be significantly different from those cases
where the correlation times of the turbulence are nearly vanishing.

A number of previous studies have exploited the idea of $\alpha$
fluctuations to study the large--scale dynamo mechanism in wide
variety of contexts. As many astrophysical
sources possess differential rotation, the focus has been on the
understanding of large--scale dynamos due to fluctuating
$\alpha$ in a shearing background
\citep{VB97,Sok97,Sil00,Pro07,Pro12,SurSub09,RP12}.
In context to the solar dynamo, \cite{Sil00} proposed a new
dynamo mechanism by considering an inhomogeneous distribution
of $\alpha$ fluctuations in a differentially rotating
atmosphere. The important difference in the present work is that
the $\alpha$ fluctuations considered here are statistically
stationary and homogeneous.
The numerical simulations of \citet{BRRK08,You08,SJ15} demonstrated
large--scale dynamo action in a shear flow with
turbulence that is, on average, non--helical.
This problem of the shear dynamo was
taken up in a number of analytical works where the
quantity $\alpha$ was assumed to be strictly zero, i.e., it
vanishes point--wise both in space and time
\citep{KR08,RK08,SS09a,SS09b,SS10,SS11}. Based on results of
\citet{SS09a,SS09b,SS10,SS11}, it was realized that the mean magnetic
field cannot grow if $\alpha$ vanishes strictly (i.e., if it
vanishes everywhere instantaneously). \citet{HMS11,McW12,MB12}
considered temporal $\alpha$ fluctuations and reported growth
of second moment of mean magnetic field, i.e., mean magnetic energy.
However, it must be noted that the possibility of the growth of
also the first moment existed in the calculations of \cite{MB12},
where the growth occurs by the process of negative diffusion.
This issue was also clarified in a recent work by \cite{SB15} who
proposed an interesting magnetic shear--current effect as being
responsible for shear dynamos.

It was realized in \cite{SS14}, and also shown in the present work,
that the negative diffusion remains the
only possibility for driving mean--field dynamos in such fluctuating
$\alpha$ calculations, so long as these fluctuations are delta--correlated
in time. On the other hand, the process of the negative diffusion is,
in a sense, self--limiting, as it would increasingly create smaller
scale structures, where the necessary assumption of the scale--separation
cannot continue to be valid indefinitely. This would
eventually lead to breakdown of the two--scale framework.
Given such limitations posed by the process of negative diffusion,
or in other words, strong $\alpha$--fluctuation dominated dynamos,
it is desirable to seek possibilities of mean--field dynamo action
when there are only weak $\alpha$ fluctuations.
\cite{SS14} found the possibility of the growth of the
mean magnetic field also in case of weak $\alpha$ fluctuations
when they considered memory effects into account.
Although many of the previous works included
shear in their studies, this highlights the
importance of fluctuations in $\alpha$, which have indeed been
measured in simulations
of \cite{BRRK08}. Estimations of such $\alpha$ fluctuations in
various astrophysical sources will be of immense value.

The aim of the present paper is to explore large--scale dynamo action, arising
solely due to an $\alpha$ that is varying stochastically in space and time, with
zero mean. We define our model in Section~2. Using the
first--order--smoothing--approximation (FOSA), we derive an integro--differential
equation governing the evolution of the large--scale magnetic field in Section~3.
This equation is non--perturbative in the $\alpha$--correlation time
$\tau_{\alpha}$. We first consider the case of white--noise $\alpha$
fluctuations (i.e., $\tau_{\alpha}=0$) without any further approximation.
Assuming small but non--zero $\alpha$--correlation time, $\tau_{\alpha}\neq 0$,
and slowly varying mean magnetic field, we simplify the integro--differential
equation to a partial differential equation (PDE) in Section~4.
Without loss of generality we explore in Section~5 one--dimensional propagating
modes and solve the dispersion relation to obtain explicit expression for the
growth rate function. In Section~6 we study dynamo action due to Kraichnan diffusivity
and Moffatt drift. We conclude in Section~7.

\section{Definition of the model}

Let us consider a fixed Cartesian coordinate system
with unit vectors $(\ex, \ey, \ez)$,
where $\bfx = (x_1, x_2, x_3)$ denotes the position vector, and
$t$ is the time variable. We model small--scale turbulence in the
absence of mean motions as random velocity fields,
$\{\bfv(\bfx, t)\}$, and consider an ensemble with its members
corresponding to different realizations of the velocity field $\bfv$.
Denoting by $\langle\;\;\rangle$ the ensemble average which obeys
standard Reynolds rules \citep[see e.g.,][]{BS05},
and assuming that the ensemble has (i) zero mean isotropic velocity
fluctuations, (ii) uniform and constant ensemble--averaged
kinetic energy density per unit mass, and
(iii) slow helicity fluctuations, we write:
\beq
\left\langle v_i\right\rangle \;=\; 0\,;\qquad
\left\langle v_i v_j\right\rangle \;=\;\delta_{ij} v_0^2 \;;\qquad
\left\langle v_i \frac{\partial v_j}{\partial x_n}\right\rangle \;=\;
\epsilon_{inj}\, \mu(\bfx, t)\,,
\eeq
\noindent
where $v_0^2 = \left\langle v^2/3\right\rangle =$~two--thirds of
the ensemble--averaged kinetic energy density per unit mass, and
$\mu(\bfx, t) = \left\langle\bfv\cendot(\bnabla\cross\bfv)
\right\rangle/6 =$ one--sixth of the ensemble--averaged helicity 
density. Let $\ell_0$ be the size of the largest eddies
and $\tau_c$ be the velocity correlation time.
By slow helicity fluctuations we
mean that the spatial and temporal scales of variation of
$\mu(\bfx, t)$ are assumed to be much larger than $\ell_0$
and $\tau_c$. 

Let $\bfB(\bfx, t)$ be the \emph{meso--scale magnetic field}, 
obtained by averaging over the above ensemble.
This requires a scale separation such that the typical scales
of $\bfB(\bfx, t)$ are much larger than $\ell_0$.
Then the space--time evolution of $\bfB(\bfx, t)$ is given by the
following equation \citep{Mof78,KR80,BS05}:
\beq
\frac{\partial \bfB}{\partial t} \;=\; 
\bnabla\cross\left[\,\alpha(\bfx, t) \bfB\,\right]\,+\,
\eta_T\bnabla^2\bfB \,;\qquad\qquad \bnabla\cendot\bfB \;=\; 0\,,
\label{DynEqn}
\eeq
\noindent
where
\begin{eqnarray}
\alpha &\;=\;& -2\tau_c\mu(\bfx, t)\,,\qquad\qquad
\eta_T \;=\; \eta \,+\,\eta_t \;=\; \mbox{total diffusivity}\,,
\nonumber\\[1ex]
\eta &\;=\;& \mbox{microscopic diffusivity}\,,\qquad\qquad
\eta_t \;=\; \tau_c v_0^2 \;=\; \mbox{turbulent diffusivity}\,.
\label{pardef}
\end{eqnarray}
\noindent
The above simple expressions for the turbulent transport coefficients
$\alpha$ and $\eta_t$ are valid only in specialized conditions ---
specifically, under the first order smoothing approximation (FOSA)
assuming isotropic turbulence, when the high--conductivity
limit is considered in conjunction with the so--called
``short--sudden'' approximation, where the velocity correlation
time $\tau_c$ is much smaller than its turnover time $\tau_0$
\citep{CHT06}.

In the ``double--averaging scheme''\footnote{The concept of
double--averaging has been discussed in a number of previous works
\citep[see e.g.,][]{Kra76,Sok97}. We refer the reader to
Section~11 of \cite{Mof83} for an excellent account of successive
averaging scheme over a number of widely separated scales, leading
to a successive renormalization of turbulent transport.}
being employed here, the helicity
fluctuations are modelled by fluctuating $\alpha$, which makes
equation~(\ref{DynEqn}) a stochastic partial differential equation.
As we are finally interested in scales much larger than the scales
of the meso--scale field, with quantity $\alpha$ being smooth around
meso scale but fluctuating at larger scales,
we can repeat the averaging procedure to obtain the evolution of
large scale field.
Here, the important step is to consider a \emph{superensemble} over which
$\alpha(\bfx, t)$ is a statistically stationary, homogeneous,
random function of $\bfx$ and $t$, with zero mean,
$\overline{\alpha(\bfx, t)}=0\,$.
The two--point space--time correlation function of fluctuating
$\alpha$:
\begin{eqnarray}
\overline{\alpha(\bfx, t) \alpha(\bfx', t')} &\;=\;& 
2{\cal A}(\bfx-\bfx')\,{\cal D}(t-t')\,,\qquad\mbox{with}
\nonumber\\[1em]
2\int_0^{\infty}\,{\cal D}(t){\rm d}t &\;=\;& 1\,,\qquad\qquad
{\cal A}({\bf 0}) \;=\; \eta_{\alpha} \,\geq\, 0\,.
\label{KMcorr}
\end{eqnarray} 
\noindent
Here $\eta_{\alpha}$ is the \emph{$\alpha$--diffusivity}, introduced
first by \citet{Kra76}. Let us define the  correlation time for the
$\alpha$ fluctuations as,
\beq
\tau_{\alpha} \;=\; 2\int_0^{\infty} \mathrm{d}t\;t\; {\cal D}(t)\,.
\label{corr-time}
\eeq
\noindent
The meso--scale field is split as
$\bfB=\overline{\bfB}+\bfb$, where $\overline{\bfB}$ is the
\emph{large--scale magnetic field} which is equal to the
superensemble--average of the meso--scale field, 
and $\bfb$ is the part of the magnetic field that fluctuates on
the meso--scales, simply referred to as the
\emph{fluctuating magnetic field} from here onwards. Applying
Reynolds averaging to the equation~(\ref{DynEqn}) and assuming
$\overline{\alpha(\bfx, t)}=0\,$, we obtain the
following equation governing the dynamics of the large--scale
magnetic field:
\begin{eqnarray}
\frac{\partial \overline{\bfB}}{\partial t} &\;=\;&
\bnabla\cross\overline{\bfemf} \;+\; \eta_T\bnabla^2\overline{\bfB}\,,
\qquad\qquad \bnabla\cendot\overline{\bfB} \;=\; 0\,,
\label{meanalpfluc}\\[2ex]
\mbox{where}\qquad\overline{\bfemf} &\;=\;&
\overline{\alpha(\bfx, t)\bfb(\bfx, t)}\,.
\label{meanEMF}
\end{eqnarray}
\noindent
To calculate $\overline{\bfemf}$, the mean \emph{electromotive force (EMF)},
we need to solve for the fluctuating field, $\bfb(\bfx, t)$, whose
evolution is determined by:
\begin{eqnarray}
\frac{\partial \bfb}{\partial t} &\;=\;&
\bnabla\cross\left[\,\alpha \overline{\bfB}\,\right] \,+\,
\bnabla\cross\left[\,\alpha\bfb-\overline{\alpha\bfb}\right]\,+\,
\eta_T\bnabla^2\bfb\,, \nonumber\\[2ex]
\bnabla\cendot\bfb &\;=\;& 0\,,
\quad \mbox{with initial condition}\quad\bfb(\bfx, 0)=\bf0\,.
\label{flucalpfluc}
\end{eqnarray}

\noindent
To keep the analysis simple while providing nontrivial extension to
the existing models, some simplifying assumptions were made, and
therefore it is useful to recall the basic limitations of our
model. Equation~(\ref{DynEqn}) assumes a local and instantaneous
relation between the meso--scale EMF and the corresponding magnetic
field. Another limitation is the choice of isotropic transport
coefficients $\alpha$ and $\eta_t$, and it is desirable to understand
the effects of fluctuations in all components of more general
tensorial $\alpha_{ij}$ and $\eta_{ij}$. This is beyond the scope
of the present investigation and will be the subject of a future study.

\section{Equation for the large--scale magnetic field}

To derive a closed equation for the large--scale magnetic field, we first
solve for the small--scale magnetic field.
Following the standard closure technique known as the FOSA,
or in other words, a quasi--linear approach, where we ignore
the mode coupling term in the fluctuating field equation, we drop the
term $\bnabla\cross\left[\,\alpha\bfb-\overline{\alpha\bfb}\right]$ from
equation~(\ref{flucalpfluc}), but retain the $\eta_T\bnabla^2\bfb$
term.\footnote{SS14 dropped the $\eta_T\bnabla^2\bfb$ term too,
for simplicity, from
the evolution equation for the fluctuating magnetic field, while studying
the shear dynamo problem. They pointed out that this would result in
overestimation of growth/decay rates for large wavenumbers. This is
confirmed later in this work.} Thus the small--scale magnetic field evolves
as:
\beq
\left(\frac{\partial}{\partial t} \;-\; \eta_T\bnabla^2 \right)\bfb \;=\;
\bnabla\cross\bfM\,,
\qquad\qquad\bnabla\cendot\bfb \;=\; 0\,,\qquad\qquad\bfb(\bfx, 0) \;=\; \bf0\,,
\label{FFE-fosa}
\eeq
\noindent
where $\bfM(\bfx, t) = \alpha(\bfx, t)\overline{\bfB}(\bfx, t)$
is a stochastic source field. Let us define the spatial Fourier transform of,
say, $\bfb(\bfx, t)$, denoted by $\widetilde{\bfb}(\bfk, t)$,
and its inverse transform as:
\beq
\widetilde{\bfb}(\bfk, t) = \int \mathrm{d}^3x\,
\exp{(-\mathrm{i}\,\bfk\cendot\bfx)}\; \bfb(\bfx, t)\;\;;
\quad \mbox{and}\quad
\bfb(\bfx, t) = \int \frac{\mathrm{d}^3k}{(2\pi)^3}\,
\exp{(\mathrm{i}\,\bfk\cendot\bfx)}\; \widetilde{\bfb}(\bfk, t)
\label{FT-IFT}
\eeq
\noindent
Fourier transforming equation~(\ref{FFE-fosa}),
we get after some algebra:
\begin{eqnarray}
\left(\frac{\partial}{\partial t} \;+\; \eta_T k^2\right) \widetilde{\bfb}
&\;=\;& \mathrm{i}\bfk\cross\widetilde{\bfM}\,,\qquad
\bfk\cendot\widetilde{\bfb} \;=\; 0\,,\qquad \widetilde{\bfb}(\bfk, 0) 
\;=\; \bf0\,,\nonumber\\[1em]
\mbox{where} \qquad
\widetilde{\bfM}(\bfk, t) &\;=\;&  
\frac{1}{(2\pi)^3} \int \mathrm{d}^3k'\,
{\widetilde{\alpha}}^{\,*}(\bfk', t)\,
\widetilde{\overline{\bfB}}(\bfk+\bfk', t)\,.
\label{FFE-fou}
\end{eqnarray}
\noindent
Solution to the equation~(\ref{FFE-fou}) satisfying the constraints,
$\bfk\cendot\widetilde{\bfb} = 0$ and
$\widetilde{\bfb}(\bfk, 0) = \bf0$, may be obtained by direct integration,
which gives
\beq
\widetilde{\bfb}(\bfk, t) \;=\; \int_0^{t} \mathrm{d}t'\,
\exp{\left[-\eta_T k^2 (t-t')\right]}\;
\left[\mathrm{i}\bfk\cross\widetilde{\bfM}(\bfk, t')\right]
\label{bsoln}
\eeq
We use equation~(\ref{bsoln}) to calculate the Fourier transform of the
mean EMF:
\begin{eqnarray}
\widetilde{\overline{\bfemf}}(\bfk, t) &=&   \int \mathrm{d}^3x
\exp{(-\mathrm{i}\,\bfk\cendot\bfx)}\,\overline{\bfemf}(\bfx, t)
=\int \mathrm{d}^3x
\exp{(-\mathrm{i}\,\bfk\cendot\bfx)}\,\overline{\alpha(\bfx, t)\,\bfb(\bfx, t)}\nonumber \\[2ex]
&=& \frac{1}{(2\pi)^3} \int \mathrm{d}^3k'\, \mathrm{d}^3k''\,
\delta(\bfk'+\bfk''-\bfk)\;\overline{\widetilde{\alpha}(\bfk', t)\,\widetilde{\bfb}(\bfk'', t)}
\nonumber \\[2ex]
&=& \frac{1}{(2\pi)^3} \int \mathrm{d}^3k'\, \mathrm{d}^3k''\,
\delta(\bfk'+\bfk''-\bfk) \nonumber \\[1ex]
&&\qquad\times
\int_0^{t} \mathrm{d}t'\,\exp{\left[-\eta_T k''^2 (t-t')\right]}
\left[\,\mathrm{i}\bfk''\cross\overline{\widetilde{\alpha}(\bfk', t)
\widetilde{\bfM}(\bfk'', t')}\,\right]
\,.\label{EMFfou} 
\end{eqnarray}
\noindent
Above expression for $\widetilde{\overline{\bfemf}}(\bfk, t)$ is given in
terms of the quantity
$\overline{\widetilde{\alpha}(\bfk', t)\widetilde{\bfM}(\bfk'', t')}\,$,
which has to be calculated. 
Using equation~(\ref{FFE-fou}) for $\widetilde{\bfM}\,$,
\beq
\overline{\widetilde{\alpha}(\bfk', t)\widetilde{\bfM}(\bfk'', t')} \;=\;
\frac{1}{(2\pi)^3} \int \mathrm{d}^3k'''\,\overline{\widetilde{\alpha}(\bfk', t)
\widetilde{\alpha}^*(\bfk''', t')}\;\widetilde{\overline{\bfB}}(\bfk''+\bfk''', t')
\label{am-eqn}
\eeq
\noindent 
is a convolution of the large--scale magnetic field and the Fourier--space $2$--point
correlator of the stochastic $\alpha$. Using equation~(\ref{KMcorr}) we write the
following expression for $2$--point correlator in Fourier space:
\begin{eqnarray}
&&\overline{\widetilde{\alpha}(\bfk', t)\widetilde{\alpha}^*(\bfk''', t')} \;=\;
\int \mathrm{d}^3x' \mathrm{d}^3x''' \exp{(-\mathrm{i}\,\bfk'\cendot\bfx'
+\mathrm{i}\,\bfk'''\cendot\bfx''')}\;\overline{\alpha(\bfx', t) \alpha(\bfx''', t')}\nonumber\\[2ex]
&&\;=\; 2{\cal D}(t-t')\int \mathrm{d}^3x' \mathrm{d}^3x'''
\exp{[-\mathrm{i}(\bfk'\cendot\bfx'-\bfk'''\cendot\bfx''')]}\,
{\cal A}\!\left(\bfx'-\bfx'''\right)\,.
\nonumber
\end{eqnarray}
\noindent
Using new integration variables, $\bfr=\bfx'-\bfx'''\,$ and 
$\bfr'=\frac{1}{2}(\bfx'+\bfx''')\,$, we get
\begin{eqnarray}
\label{aa-corr}
\overline{\widetilde{\alpha}(\bfk', t)\widetilde{\alpha}^*(\bfk''', t')} &=&
2{\cal D}(t-t')\int \mathrm{d}^3r\, \mathrm{d}^3r'\;
\exp{\left[\,-\mathrm{i}(\bfk'-\bfk''')
\cendot\bfr' - \frac{\mathrm{i}}{2}(\bfk'+\bfk''')\cendot\bfr\,\right]}
\;{\cal A}\!\left(\bfr\right)\nonumber\\[2ex]
&=& 2{\cal D}(t-t')(2\pi)^3\, \delta(\bfk'-\bfk''')
\widetilde{{\cal A}}(\bfk') \\[2ex]
\mbox{where} \qquad\qquad
\widetilde{{\cal A}}(\bfk) &=& \int \mathrm{d}^3r\,
\exp{(-\mathrm{i}\,\bfk\cendot\bfr)}\; {\cal A}(\bfr)\,.
\end{eqnarray}
\noindent
$\widetilde{{\cal A}}(\bfk)$ is the \mbox{complex spatial power spectrum}
of $\alpha$ fluctuations, with
$\widetilde{{\cal A}}(-\bfk) = \widetilde{{\cal A}}^*(\bfk)$ because
${\cal A}(\bfr)$ is a real function.
From equations~(\ref{am-eqn}) and (\ref{aa-corr}) we write
\beq
\overline{\widetilde{\alpha}(\bfk', t)\widetilde{\bfM}(\bfk'', t')} \;=\;
2{\cal D}(t-t')\widetilde{{\cal A}}(\bfk')\,
\widetilde{\overline{\bfB}}(\bfk'+\bfk'', t')\,.
\label{am-fin}
\eeq
\noindent
When equation~(\ref{am-fin}) is substituted in (\ref{EMFfou}) we obtain a 
compact expression for the EMF:
\begin{eqnarray}
\widetilde{\overline{\bfemf}}(\bfk, t) &=& 2\int_0^{t} \mathrm{d}s\,{\cal D}(s)
\left\{\,\widetilde{\bfU}(\bfk, s)\cross\widetilde{\overline{\bfB}}(\bfk, t-s)\,\right\}\,,
\label{EMF-U} \\[2ex]
\mbox{where}\qquad
\widetilde{\bfU}(\bfk, s) &=&
\int\frac{\mathrm{d}^3k'}{(2\pi)^3}\,
\exp{\left[-\eta_T (\bfk-\bfk')^2 s\right]}\,
\mathrm{i}(\bfk-\bfk') \widetilde{{\cal A}}(\bfk')\,.
\label{Ugen}
\end{eqnarray}
\noindent
Fourier transforming equation~(\ref{meanalpfluc}), the equation governing 
the large--scale field is:
\beq
\frac{\partial \widetilde{\overline{\bfB}}}{\partial t} \;=\; 
\mathrm{i}\bfk\cross\widetilde{\overline{\bfemf}} \;-\; 
\eta_T k^2\,\widetilde{\overline{\bfB}}\,,\qquad\qquad
\bfk\cendot \widetilde{\overline{\bfB}} \;=\; 0\,.
\label{MFE-fou}
\eeq

\noindent
Thus the set of equations~(\ref{EMF-U})--(\ref{MFE-fou}) describe the evolution of
the large--scale magnetic field, $\widetilde{\overline{\bfB}}$. Solving these in full
generality is beyond the scope of the present investigation, and we study this system
of closed equations analytically in useful approximations.

\subsection{White--noise $\alpha$ fluctuations}

Before studying finite memory effects, we consider the exactly solvable limit
of white--noise (i.e., delta--correlated--in--time) $\alpha$ fluctuations,
for which, the normalized correlation function
is ${\cal D}_{\!{\rm WN}}(t) = \delta(t)\,$, the Dirac delta--function.
This gives $\tau_{\alpha}=0$ from equation~(\ref{corr-time}), implying that
the memory effects are ignored.
The focus of this section is to understand the dynamo behaviour due to
white--noise $\alpha$ fluctuations. From equation~(\ref{EMF-U}) the mean EMF
for white--noise is
\beq
\widetilde{\overline{\bfemf}}_{\rm WN}(\bfk, t) 
\;=\; \widetilde{\bfU}(\bfk, 0)\cross\widetilde{\overline{\bfB}}(\bfk, t)\,,
\label{EMF-wn-fou}
\eeq
\noindent
which, as expected, depends on the large--scale field at the present time only.
The quantity $\widetilde{\bfU}(\bfk, 0)$ may be simplified as:
\begin{eqnarray}
\widetilde{\bfU}(\bfk, 0) &\;=\;& 
\mathrm{i}\bfk\int\frac{\mathrm{d}^3k'}{(2\pi)^3}\,
\widetilde{{\cal A}}(\bfk') \;-\; 
\int\frac{\mathrm{d}^3k'}{(2\pi)^3}\,\mathrm{i}\bfk'
\widetilde{{\cal A}}(\bfk')
\nonumber\\[1em]
&\;=\;& \mathrm{i}\bfk{\cal A}({\bf 0}) \;-\; 
\frac{\partial{\cal A}(\bfxi)}{\partial\bfxi}{\Biggl {\vert}}_{\bfxi = {\bf 0}}
\;\;\;=\;\; \mathrm{i}\bfk \eta_{\alpha} \;+\; \bfV_{\!\!M}\,,
\label{Uwn}
\end{eqnarray}
Here, $\eta_{\alpha}$ is the $\alpha$--diffusivity defined in equation~(\ref{KMcorr}),
and $\bfV_{\!\!M}$ is the Moffatt drift velocity \citep{Mof78,SS14}.
Both, $\eta_{\alpha}$ and $\bfV_{\!\!M}$, are constants by definition.
Substituting (\ref{EMF-wn-fou}) in (\ref{MFE-fou}) and using equation~(\ref{Uwn}),
we get the following partial differential equation for the large--scale magnetic field:
\beq
\frac{\partial \widetilde{\overline{\bfB}}}{\partial t} \;+\;
\left[\,\eta_K k^2 \,+\, \mathrm{i}\, \bfk\cendot\bfV_{\!\!M}\,\right]\,
\widetilde{\overline{\bfB}} \;=\; {\bf 0}\,,
\qquad\qquad \bfk\cendot\widetilde{\overline{\bfB}} \;=\; 0\,.
\label{MFE-wn}
\eeq
\noindent
Here
\begin{eqnarray}
\eta_K &\;=\;& \eta_T \,-\, \eta_{\alpha} \;=\;
\mbox{Kraichnan diffusivity}\,,
\nonumber\\[1em] 
\bfV_{\!\!M} &\;=\;& -\left(\frac{\partial {\cal A}(\bfxi)}
{\partial\bfxi}\right)_{\bfxi =\bf0} \;=\; \int_0^\infty
\overline{\alpha(\bfx, t)\bnabla\!\alpha(\bfx, 0)}\,{\rm d}t 
\;=\; \mbox{Moffatt drift velocity}\,, 
\label{kmcondef}
\end{eqnarray}
\noindent
are the two constants that determine the behaviour of the
large--scale magnetic field. Note that the $\alpha$ diffusivity
contributes a decrement to the diffusivity, and hence aids dynamo
action \citep{Kra76,Mof78,SS14}.
The solution to equation~(\ref{MFE-wn}) is given by
\begin{eqnarray}
\widetilde{\overline{\bfB}}(\bfk, t) &=&
\widetilde{{\cal G}}(\bfk, t)\,\widetilde{\overline{\bfB}}(\bfk, 0)\,,
\qquad\qquad \bfk\cendot \widetilde{\overline{\bfB}} \;=\; 0\,,
\label{BG} \\[2ex]
\mbox{where}\qquad
\widetilde{{\cal G}}(\bfk, t) &=& \exp{\Bigl\{-\eta_K k^2t \,-\,
\mathrm{i}\left(\bfV_{\!\!M}\cendot\bfk\right) t\Bigr\}}\,.
\label{greenfn}
\end{eqnarray}
\noindent
Equations~(\ref{BG}) and (\ref{greenfn}) provide complete solution to the
problem of white--noise $\alpha$ fluctuations, where the growth or decay
of mean magnetic field $\widetilde{\overline{\bfB}}$ is
determined by the Green's function $\widetilde{{\cal G}}(\bfk, t)$.
We note some general properties:
\begin{itemize}
\item[1.] {\bf Weak $\alpha$ fluctuations} have $\eta_{\alpha} < \eta_T\,$ 
so that $\eta_K > 0\,$. In this case modes of all wavenumbers $\bfk$ decay. 
\item[2.] {\bf Strong $\alpha$ fluctuations} have $\eta_{\alpha} > \eta_T\,$
so that $\eta_K < 0\,$. This belongs to the case when $\alpha$--diffusivity
compensates and overcomes the total (turbulent+microscopic)
diffusivity $\eta_T$.
This leads to the growth of modes of all wavenumbers $\bfk$ by the
process of \emph{negative diffusion} which was first obtained by \cite{Kra76}.
It may be noted that the process of the negative diffusion is,
in a sense, self--limiting, as it would increasingly create smaller
scale structures, where the necessary assumption of the scale--separation
cannot continue to be valid indefinitely. This would
eventually lead to breakdown of the two--scale framework.
\item[3.] The Moffatt drift velocity $\bfV_{\!\!M}$ contributes only to
the phase and does not determine the growth or decay of
the large--scale magnetic field.
\end{itemize}
\emph{Therefore, the necessary condition for dynamo action for white--noise
$\alpha$ fluctuations is that they must be strong, i.e., $\eta_K<0$.}
It is necessary to consider
${\cal D}(t)\neq \delta(t)$ to explore memory effects, which will have
$\tau_{\alpha}\neq 0$. This is studied in the next section.

\section{Large--scale magnetic fields when $\tau_{\alpha}$ is small}

Now we extend our analysis to include the effects of finite memory of
fluctuating $\alpha$.
By assuming small $\tau_{\alpha}$, we reduce general integro--differential
equation, given by equations~(\ref{EMF-U})--(\ref{MFE-fou}), to a PDE governing
the evolution of large--scale magnetic field which evolves over times much
larger than $\tau_{\alpha}$.
Then we first consider the Kraichnan problem with nonzero $\tau_{\alpha}$,
but without Moffatt drift $\bfV_{\!\!M}$, which is studied in detail
in Section~6.

\subsection{Derivation of the governing equation}

We note that the normalized time correlation function, ${\cal D}(t)$,
has a singular limit:  i.e.
$\lim\limits_{\tau_{\alpha}\to 0}{\cal D}(t)={\cal D}_{\rm WN}(t)=\delta(t)$. 
Here we wish to consider non--zero but small $\tau_{\alpha}$ which implies
that the function ${\cal D}(t)$ is significant only for times
$t\leq \tau_{\alpha}$ and becomes negligible for larger times.
We simplify the mean EMF given by equation~(\ref{EMF-U}), together with (\ref{Ugen}),
by solving the time integral for small $\tau_{\alpha}$.
Since the limit $\lim\limits_{\tau_{\alpha}\to 0}
\widetilde{\overline{\bfemf}}(\bfk, t) = 
\widetilde{\overline{\bfemf}}_{\rm WN}(\bfk, t) = 
\widetilde{\bfU}(\bfk, 0)\cross\widetilde{\overline{\bfB}}(\bfk, t)$
is evidently non singular, we make the \emph{ansatz} that,
for small $\tau_\alpha$, the EMF can be expanded in a power series in 
$\tau_{\alpha}$:
\beq
\widetilde{\overline{\bfemf}}(\bfk, t) \;=\; \widetilde{\overline{\bfemf}}_{\rm WN}(\bfk, t)
\,+\, \widetilde{\overline{\bfemf}}^{(1)}(\bfk, t) \,+\,
\widetilde{\overline{\bfemf}}^{(2)}(\bfk, t) \,+\, \,\dots
\label{Eexp}
\eeq
\noindent
where $\widetilde{\overline{\bfemf}}_{\rm WN}(\bfk, t) \sim {\cal O}(1)$
and $\widetilde{\overline{\bfemf}}^{(n)}(\bfk, t) \sim {\cal O}(\tau_{\alpha}^n)$ for $n \geq 1$.
Below we verify this ansatz up to $n= 1$, for slowly varying magnetic fields.

We wish to determine $\widetilde{\overline{\bfemf}}(\bfk, t)$ which is correct to
first order in $\tau_{\alpha}\,$, for $t\gg\tau_{\alpha}\,$.
Since ${\cal D}(s)$ is strongly peaked for times
$s\leq \tau_{\alpha}$ and becomes negligible for larger $s$ as mentioned above,
most of the contribution to the integral in (\ref{EMF-U}) comes only
from short times $0\leq s < \tau_{\alpha}$. Hence in (\ref{EMF-U}) we can
(i) set the upper limit of the $s$--integral to $+\infty\,$; (ii) keep the terms
inside the $\{\;\}$ in the integrand up to only first order in $s$.
Below we first work out $\widetilde{\bfU}(\bfk, s)$ and
$\widetilde{\overline{\bfB}}(\bfk, t-s)$ correct upto ${\cal O}(s)$.

\noindent
$\bullet\;\;\underline{\widetilde{\bfU}(\bfk, s)\; \mbox{to}\; {\cal O}(s) :}$
Taylor expansion of the function $\widetilde{\bfU}(\bfk, s)$ gives,
\beq
\widetilde{\bfU}(\bfk, s) \;=\; \widetilde{\bfU}(\bfk, 0) \,+\,
s \frac{\partial \widetilde{\bfU}}{\partial s}{\Biggl {\vert}}_{s = 0}
\,+\, {\cal O}(s^2)\,.
\eeq
\noindent
Let us rewrite $\widetilde{\bfU}(\bfk, s)$, correct upto ${\cal O}(s)$, as,
\begin{eqnarray}
 \widetilde{\bfU}(\bfk, s) &\;=\;& \bfP(\bfk)\,+\, s\,\bfQ(\bfk) \\[2ex]
 \mbox{where}\qquad \bfP(\bfk) &\;=\;& \widetilde{\bfU}(\bfk, 0) \;=\;
 \mathrm{i}\bfk \eta_{\alpha} \;+\; \bfV_{\!\!M}
 \qquad\mbox{(from equation~(\ref{Uwn}))}\,, \label{Pdef}\\[2ex]
 \mbox{and}\qquad \bfQ(\bfk) &\;=\;& \frac{\partial \widetilde{\bfU}}{\partial s}
 {\Biggl {\vert}}_{s = 0} \;=\; -\mathrm{i} \eta_T
 \int \frac{\mathrm{d}^3k'}{(2\pi)^3}\, (\bfk-\bfk')^2 (\bfk-\bfk')
 \widetilde{{\cal A}}(\bfk')\,.\label{Qdef}
\end{eqnarray}
\noindent
The integrand in equation~(\ref{Qdef}) can be expanded to obtain the following
expression for $\bfQ(\bfk)$ (see the appendix)\,:
\beq
\bfQ(\bfk) \,=\, -\eta_T k^2 \left(\mathrm{i}\bfk \eta_{\alpha}
\,+\, \bfV_{\!\!M}\right) \,-\, 2\eta_T (\bfk\,\cendot \bfV_{\!\!M}) \bfk
\,+\, \mathrm{i}\bfk \eta_T C_1 \,+\, 2\mathrm{i}\eta_T
\bfk\,\cendot \overleftrightarrow{\bfC_2} \,-\,
\eta_T \bfC_3\,,
\label{Q2}
\eeq
\noindent
where $\eta_{\alpha}$ and $\bfV_{\!\!M}$ are the constants defined in
equations~(\ref{KMcorr}) and (\ref{kmcondef}), respectively,
which shows that they depend on the spatial
correlation function ${\cal A}$ and its first spatial derivative.
Both these constants appeared already in \cite{SS14}.
But $C_1$ (scalar), $\overleftrightarrow{\bfC_2}$ (dyad) and $\bfC_3$ (vector)
are three new constants given by (see the appendix),
\beq
C_1 \,=\, \left[\nabla^2 {\cal A}(\bfxi)\right]_{\bfxi\,=\,{\bf 0}}\;;\quad
\overleftrightarrow{\bfC_2} \,=\,
\left[\bnabla \bnabla {\cal A}(\bfxi)\right]_{\bfxi\,=\,{\bf 0}}\;;\quad
\bfC_3 \,=\, \left[\nabla^2 \{\bnabla {\cal A}(\bfxi)\}\right]_{\bfxi\,=\,{\bf 0}}\,,
\label{C}
\eeq
\noindent
which depend on second or third order spatial derivatives of ${\cal A}$. Thus,
in general, the mean EMF will be determined by five constants
$(\eta_K,\,\bfV_{\!\!M},\,C_1,\,\overleftrightarrow{\bfC_2},\,\bfC_3)$ which
can all be explicitly found once the form of spatial correlation function
${\cal A}(\bfxi)$ is chosen. To keep the analysis simple, we assume that the
function ${\cal A}(\bfxi)$ is such that its second or higher order spatial
derivatives are negligible at $\bfxi={\bf 0}$, and therefore we ignore the constants
$C_1$, $\overleftrightarrow{\bfC_2}$ and $\bfC_3$ in the present work.
Thus we write:
\beq
\bfQ(\bfk) \,=\, -\eta_T k^2 \bfP(\bfk) \,-\,
2\eta_T (\bfk\,\cendot \bfV_{\!\!M}) \bfk\,.
\label{Qused}
\eeq

\noindent
$\bullet\;\;\underline{\widetilde{\overline{\bfB}}(\bfk, t-s)\; \mbox{to}\; {\cal O}(s) :}$
Taylor expansion of the function $\widetilde{\overline{\bfB}}(\bfk, t-s)$ gives,
\beq
\widetilde{\overline{\bfB}}(\bfk, t-s) \;=\; \widetilde{\overline{\bfB}}(\bfk, t)
\,-\,s \frac{\partial \widetilde{\overline{\bfB}}(\bfk, t)}{\partial t} \,+\, \ldots\,.
\label{Taylor-B}
\eeq
\noindent
where it is assumed that $\bigl{\vert} \widetilde{\overline{\bfB}}
\bigr{\vert} \,\gg\, s\bigl{\vert}\partial \widetilde{\overline{\bfB}}/
\partial t\bigr{\vert} \,\gg\, s^2\bigl{\vert}\partial^2 
\widetilde{\overline{\bfB}}/\partial t^2\bigr{\vert}\,\gg \,
s^3\bigl{\vert}\partial^3 \widetilde{\overline{\bfB}}/\partial 
t^3\bigr{\vert}\,\mbox{etc}\,$.
In order to find $\widetilde{\overline{\bfB}}(\bfk, t-s)$ which is correct upto
${\cal O}(s)$, we need $\partial \widetilde{\overline{\bfB}}/\partial t$ in
equation~(\ref{Taylor-B}) only upto ${\cal O}(1)$.
Using equation~(\ref{Eexp}) together with (\ref{EMF-wn-fou}) and (\ref{Uwn})
in equation~(\ref{MFE-fou}) we find,
\beq
\frac{\partial \widetilde{\overline{\bfB}}}{\partial t}\Biggr{\vert}_{{\cal O}(1)} \;=\;
\mathrm{i}\bfk\cross\widetilde{\overline{\bfemf}}_{\rm WN}(\bfk, t) \;-\;
\eta_Tk^2 \widetilde{\overline{\bfB}} 
\;=\; -\left[\,\mathrm{i}\bfk\,\cendot\bfP \,+\, \eta_Tk^2\right]
\widetilde{\overline{\bfB}} \,,
\label{Bdot-0thord}
\eeq
\noindent
which is substituted in equation~(\ref{Taylor-B}) to obtain:
\beq
\widetilde{\overline{\bfB}}(\bfk, t-s) \;=\; 
\widetilde{\overline{\bfB}}(\bfk, t)\left[\,1 \,+\,
s\left(\mathrm{i}\bfk\,\cendot\bfP \,+\, \eta_Tk^2\right)
\,\right] \,+\, {\cal O}(s^2)
\eeq
\noindent
Then, from the above analysis, we may write,
\begin{eqnarray}
&&\widetilde{\bfU}(\bfk, s)\cross\widetilde{\overline{\bfB}}(\bfk, t-s)=
\bfP(\bfk)\cross\widetilde{\overline{\bfB}}(\bfk, t)\;+\nonumber \\[2ex]
&&\qquad\qquad+\;s\left[\left(\mathrm{i}\bfk\,\cendot\bfP \,+\, \eta_Tk^2\right)
\bfP(\bfk)\cross\widetilde{\overline{\bfB}}(\bfk, t) \,+\,
\bfQ(\bfk)\cross\widetilde{\overline{\bfB}}(\bfk, t) \right]
+ {\cal O}(s^2)\,.
\end{eqnarray}
\noindent
Noting that $\widetilde{\overline{\bfemf}}_{\rm WN}(\bfk, t)=
\bfP(\bfk)\cross\widetilde{\overline{\bfB}}(\bfk, t)$ and using the definition of
$\bfQ(\bfk)$ as given in equation~(\ref{Qused}), we get:
\beq
\{\;\}\;\mbox{of eqn.~(\ref{EMF-U})} \;=\;  
\widetilde{\overline{\bfemf}}_{\rm WN}(\bfk, t) \,+\, 
s\left\{(\mathrm{i}\bfk\,\cendot\bfP)\,
\widetilde{\overline{\bfemf}}_{\rm WN} \,-\,
2\eta_T(\bfk\,\cendot \bfV_{\!\!M})\,\bfk\cross
\widetilde{\overline{\bfB}}(\bfk, t)
\right\} + {\cal O}(s^2)\,.
\label{integrand}
\eeq
\noindent
Using equation~(\ref{integrand}) and the properties of ${\cal D}(t)$,
given in (\ref{KMcorr}) and (\ref{corr-time}), we solve the integral
over $s$ in equation~(\ref{EMF-U}) to obtain the mean EMF:
\beq
\widetilde{\overline{\bfemf}}(\bfk, t) \;=\; \widetilde{\overline{\bfemf}}_{\rm WN}(\bfk, t)
\,+\, \tau_{\alpha}\left\{(\mathrm{i}\bfk\,\cendot\bfP)\,
\widetilde{\overline{\bfemf}}_{\rm WN} \,-\,
2\eta_T(\bfk\,\cendot \bfV_{\!\!M})\,\bfk\cross
\widetilde{\overline{\bfB}}(\bfk, t)
\right\}
\label{EMF-taualp}
\eeq
\noindent
accurate to ${\cal O}(\tau_{\alpha})$, which verifies the ansatz of
equation~(\ref{Eexp}) up to $n = 1$, as claimed. 
We note that equation~(\ref{EMF-taualp}) for the mean
EMF is valid only when the large--scale magnetic field is slowly varying.
To lowest order this condition can be stated as:
$\,\bigl{\vert} \widetilde{\overline{\bfB}}
\bigr{\vert} \gg \tau_{\alpha}\bigl{\vert}\partial \widetilde{\overline{\bfB}}/\partial t
\bigr{\vert}\,$. Using equation~(\ref{Bdot-0thord})
for $\,\partial \widetilde{\overline{\bfB}}/\partial t \,$, we see that the
sufficient condition for equation~(\ref{EMF-taualp}) to be valid is that the following two 
dimensionless quantities be small:
\beq
\vert \eta_K k^2 \tau_{\alpha}\vert \,\ll\, 1\,,\qquad\qquad
\vert k V_M \tau_{\alpha}\vert \,\ll\,1\,.
\label{cond}
\eeq
\noindent
Substituting (\ref{EMF-taualp}) in (\ref{MFE-fou}) we obtain the following
PDE governing the evolution of large--scale magnetic field: 
\beq
\frac{\partial \widetilde{\overline{\bfB}}}{\partial t} \,=\,
\biggl\{
\left[\,-\, \eta_K k^2 \,-\, \eta_{\alpha}^2 k^4\tau_{\alpha}
\,+\, (\bfk\cendot\bfV_{\!\!M})^2 \tau_{\alpha}\,\right] \,+\,
\mathrm{i}\,(\bfk\cendot\bfV_{\!\!M})
\left[\,2(\eta_{\alpha}+\eta_T) k^2 \tau_{\alpha} \,-\, 1\,\right]
\biggr\}\,\widetilde{\overline{\bfB}}
\label{MFE-smalltaualp}
\eeq

\subsection{The Kraichnan problem with non zero $\tau_\alpha$}

First we consider the Kraichnan problem and extend it to include finite
$\tau_\alpha$ in order to understand the combined effect of the $\alpha$
fluctuations when $\bfV_{\!\!M}= {\bf 0}$, but $\eta_{\alpha}>0$ and
$\tau_{\alpha}>0$. SS14 defined a length scale whose corresponding
wavenumber,
\beq
k_\alpha \;=\; \left(\eta_\alpha\tau_\alpha\right)^{-1/2} \,\;>\;\, 0\,.
\label{kalpha}
\eeq
This is used to define high or low wavenumbers ($k$);
$\vert k\vert > k_\alpha$ are called high wavenumbers, and
$\vert k\vert < k_\alpha$ are called low wavenumbers.
When $\bfV_{\!\!M}= {\bf 0}\,$, one of the two conditions
in (\ref{cond}) are met trivially, and the other one implies that 
$\vert k\vert$ must be small enough such that
$\vert\eta_K k^2\tau_\alpha\vert \ll 1\,$. Using (\ref{kalpha}) and
setting $\bfV_{\!\!M}= {\bf 0}$ in equation~(\ref{MFE-smalltaualp}), we get
\beq
\frac{\partial \widetilde{\overline{\bfB}}}{\partial t} \;=\;
-\eta_K k^2\left[ 1 \,+\, \frac{\eta_{\alpha}}{\eta_K}
\left(\frac{k}{k_\alpha}\right)^2\right]\,\widetilde{\overline{\bfB}}\,,
\qquad\mbox{with}\qquad \bfk\,\cendot\widetilde{\overline{\bfB}}\;=\; 0\,.
\label{krcase}
\eeq
\begin{figure}
\begin{center}
\includegraphics[width=\columnwidth]{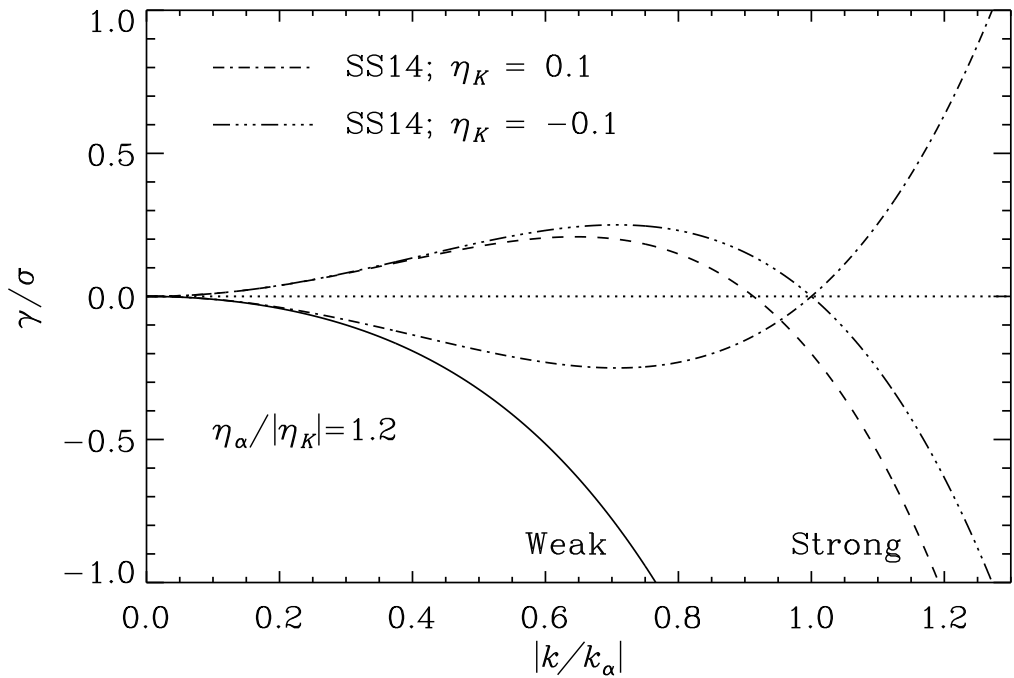}
\end{center}
\caption[]{Growth rate $\gamma/\sigma$ as a function of
$\vert k/k_{\alpha}\vert\,$, when $\bfV_{\!\!M}= {\bf 0}\,$.
Weak ($\eta_K>0$) and strong ($\eta_K<0$)
$\alpha$ fluctuations correspond to bold and dashed curves, respectively.
Results of SS14 are also shown; Dash--dotted and
triple--dot--dashed curves correspond to weak and strong $\alpha$
fluctuations, respectively.}
\label{Fig1}
\end{figure}
\noindent
The solutions are of the exponential form, 
$\widetilde{\overline{\bfB}}(\bfk, t) =
\widetilde{\overline{\bfB}}_0(\bfk)\exp{(\gamma t)}$, where
$\bfk\,\cendot\widetilde{\overline{\bfB}}_0(\bfk) = 0$. Substituting
this in equation~(\ref{krcase}), we get the growth rate as:
\beq
\gamma \;=\; -\left(\eta_K k_\alpha^2\right)
\left(\frac{k}{k_\alpha}\right)^2 \left[1 \,+\,
\frac{\eta_{\alpha}}{\eta_K}\left(\frac{k}{k_\alpha}\right)^2\right]\,,
\qquad\mbox{when}\qquad \vert\eta_K k^2\tau_\alpha\vert \ll 1\,.
\label{gammakr}
\eeq
\noindent
Following SS14 we normalize the growth rates by the characteristic frequency
$(\sigma)$ which is defined as
\beq
\sigma \;=\; \vert\eta_K\vert k_\alpha^2
\;=\; \left(\frac{\vert\eta_T - \eta_\alpha\vert}{\eta_\alpha}\right)\frac{1}{\tau_\alpha}
\;\geq\; 0\,.
\label{sigma}
\eeq
\noindent
Below we consider the behaviour of the growth rate as a function of the wavenumber,
for weak and strong $\alpha$ fluctuations, and show this in Fig.~(\ref{Fig1}) where
we also show the corresponding results of SS14 to illustrate some qualitative
differences.\footnote{We note, however, that the
expressions for the growth rates in SS14 were simpler and did not
explicitly depend on the factor $\eta_{\alpha}/|\eta_K|$.}

\noindent
{\bf Weak $\alpha$ fluctuations:} This has $0 < \eta_\alpha < \eta_T$, 
so that $\eta_K$ is positive. The normalized growth rate ($\gamma/\sigma$)
may be expressed as
\beq
\frac{\gamma}{\sigma} \;=\; -\left(\frac{k}{k_\alpha}\right)^2
\left[1 \,+\, \frac{\eta_{\alpha}}{\vert\eta_K\vert}
\left(\frac{k}{k_\alpha}\right)^2\right]\,,\qquad\mbox{when}\qquad
\vert k\vert \ll \frac{k_{\alpha}}{\sqrt{\sigma \tau_{\alpha}}}
\label{gammakr-w}
\eeq
\noindent
The growth rate is negative definite for finite $k$ and it is a monotonically
decreasing function of the wavenumber; shown by bold curve in Fig.~(\ref{Fig1}).
This is even qualitatively different from SS14 where the high wavenumbers
always grow for weak $\alpha$ fluctuations (dash--dotted curve).
This also highlights the fact that the inclusion of the resistive term in
the fluctuating field equation is a nontrivial extension of the SS14
model.

\noindent
{\bf Strong $\alpha$ fluctuations:} This has $0 < \eta_T < \eta_\alpha$, 
so that $\eta_K$ is negative. The normalized growth rate ($\gamma/\sigma$)
may be expressed as
\beq
\frac{\gamma}{\sigma} \;=\; +\left(\frac{k}{k_\alpha}\right)^2
\left[1 \,-\, \frac{\eta_{\alpha}}{\vert\eta_K\vert}
\left(\frac{k}{k_\alpha}\right)^2\right]\,,\qquad\mbox{when}\qquad
\vert k\vert \ll \frac{k_{\alpha}}{\sqrt{\sigma \tau_{\alpha}}}
\label{gammakr-s}
\eeq
In this case the growth rate can be positive for a range of wavenumbers,
before becoming negative at larger wavenumbers; shown by bold curve in
Fig.~(\ref{Fig1}). This is qualitatively similar to the results of SS14
(triple--dot--dashed curve) in this regime and the differences at large
wavenumbers arise due to reasons mentioned above.

\emph{Thus the necessary condition for dynamo action is that the $\alpha$
fluctuations must be strong, i.e., $\eta_K<0$. Recall that this is, in a sense,
similar to the white--noise or the original Kraichnan model, but we note the
following important difference: in white--noise case, the growth rate increases
monotonically with wavenumber ($k$) for strong $\alpha$ fluctuations,
with largest allowed wavenumbers
(smallest allowed length scales) growing the fastest; whereas, here, we find
that the growth rate is a nonmonotonic function of $k$, and as a result,
there exists a wavenumber cutoff at some large $k$ beyond which the growth
rate turns negative.}
This makes it a special dynamo as the magnetic power at
smallest length scales would be suppressed due to the existence
of wavenumber cutoff
(see dashed curve in Fig.~(\ref{Fig1})), thus enabling a bonafide large--scale
dynamo, unlike the white--noise case where much of the magnetic power lies at
the smallest allowed length scales.

It would be indeed interesting if even weak $\alpha$ fluctuations could
lead to large--scale dynamo action.
In the next section we explore the combined effect of $\bfV_{\!\!M}\neq0$
and $\tau_{\alpha}\neq0$, and ask the following question:
\emph{when both $\tau_{\alpha}$ and $\bfV_{\!\!M}$ are nonzero,
could large--scale magnetic fields grow even when $\alpha$ fluctuations
are weak, i.e. when $\eta_K > 0\,$}?

\section{Growth rates of modes when $\tau_\alpha$ is non zero}

We consider one--dimensional propagating modes for the general case
when all the parameters $(\eta_\alpha, \bfV_{\!\!M}, \tau_\alpha)$
can be non zero. Below we derive the dispersion relation and study
the growth rate function.  When the wavevector $\bfk = (0,0,k)$ points
along the ``vertical'' ($\pm\ez$) directions,
$\widetilde{\overline{B}}_3$ must be uniform and is of no interest
for dynamo action. Hence we set $\widetilde{\overline{B}}_3 = 0$, and
take $\widetilde{\overline{\bfB}}(k, t) = \widetilde{\overline{B}}_1(k, t)
\ex \,+\, \widetilde{\overline{B}}_2(k, t)\ey\,$. The equation governing
the time evolution of this large--scale magnetic field is obtained by
setting $k_{1,2} = 0$, $\,k_3=k\,$ and $\,\widetilde{\overline{B}}_3 =0\,$
in equation~(\ref{MFE-smalltaualp}):
\beq
\frac{\partial \widetilde{\overline{\bfB}}}{\partial t} \,=\,
\biggl\{
\left[\,-\, \eta_K k^2 \,-\, \eta_{\alpha}^2 k^4\tau_{\alpha}
\,+\, (k V_{\!M3})^2 \tau_{\alpha}\,\right] \,+\,
\mathrm{i}\,k V_{\!M3}
\left[\,2(\eta_{\alpha}+\eta_T) k^2 \tau_{\alpha} \,-\, 1\,\right]
\biggr\}\,\widetilde{\overline{\bfB}}
\label{Beq1d}
\eeq
\noindent
We note that each component of mean magnetic field evolves independently
of other components. The nature of this dynamo is therefore different
from standard $\alpha^2$- or $\alpha \omega$-dynamos where evolutions of
various components are coupled with each other thus facilitating
cross--coupling dynamo.
We seek modal solutions of the form,
\beq
\widetilde{\overline{\bfB}}(k, t) \;=\; 
\left[\widetilde{\overline{B}}_{01}(k)\ex \,+\, 
\widetilde{\overline{B}}_{02}(k)\ey\right]\,\exp{(\lambda t)}\,,
\label{modal}  
\eeq
\noindent
and substitute it in equation~(\ref{Beq1d}) to obtain the following
\emph{dispersion relation}:
\beq
\lambda \;=\; \left[\,-\, \eta_K k^2 \,-\, \eta_{\alpha}^2 k^4\tau_{\alpha}
\,+\, (k V_{\!M3})^2 \tau_{\alpha}\,\right] \,+\,
\mathrm{i}\,k V_{\!M3}
\left[\,2(\eta_{\alpha}+\eta_T) k^2 \tau_{\alpha} \,-\, 1\,\right]
\label{disprel}
\eeq
\noindent
Of particular interest is the growth rate $\gamma = {\rm Re}\{\lambda\}\,$,
because dynamo action corresponds to the case when $\gamma > 0\,$. From 
the dispersion relation~(\ref{disprel}) we have:
\beq
\gamma \;=\; -\, \eta_K k^2 \,-\, \eta_{\alpha}^2 k^4\tau_{\alpha}
\,+\, (k V_{\!M3})^2 \tau_{\alpha}
\label{gamma}
\eeq
We refer the reader to Appendix~B for some properties of the growth rate
function in terms of useful dimensionless parameters and to Appendix~C
for their physical meanings.

\section{Dynamo action due to Kraichnan diffusivity and Moffatt drift}

Now we turn to the most general case when both, Kraichnan diffusivity
and Moffatt drift, are nonzero, and $\alpha$--fluctuations have finite
correlation times. SS14 defined a new time--scale involving
$\eta_K$ and $V_{\!M3}$ as:
\beq
\tau_{*} \;=\; (\vert\eta_K\vert/V_{\!M3}^2) \;>\; 0\,.
\label{taustar}
\eeq
We provide below the expressions for dimensional growth rate $\gamma$
as a function of the wavenumber $k$, for weak and strong $\alpha$
fluctuations. It turns out that the nature of dynamo action depends
on whether $\tau_{\alpha}$ is smaller or larger than $\tau_{*}$.

\begin{figure}
\begin{center}
\includegraphics[width=\columnwidth]{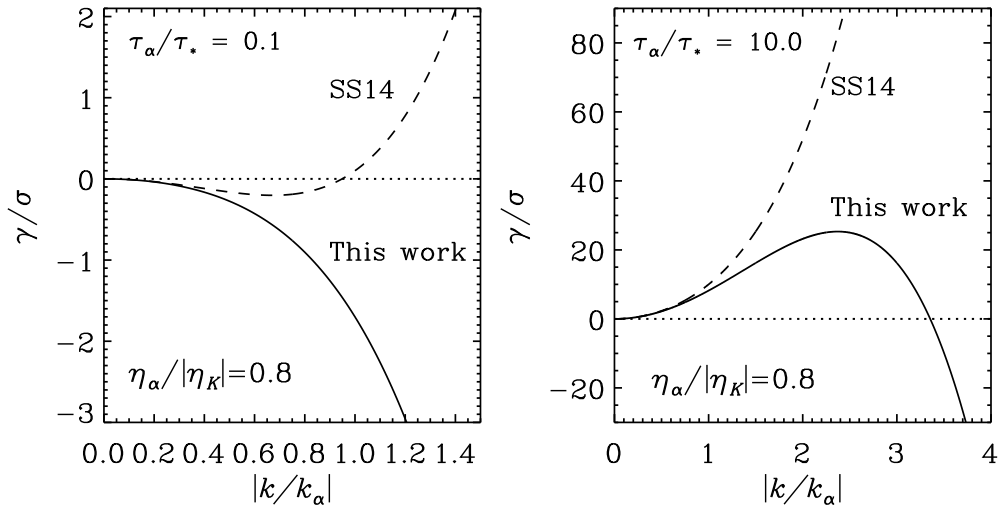}
\end{center}
\caption[]{Growth rate $\gamma/\sigma$ plotted as a function of
$\vert k/k_{\alpha}\vert$ for \emph{weak $\alpha$ fluctuations};
shown by bold curves. Panels (a) and (b) correspond to the case when
$\tau_{\alpha}/\tau_* = 0.1$ and $10.0$, respectively.
Dashed curves correspond to the results of SS14.}
\label{Fig3}
\end{figure}

\noindent
{\bf Weak $\alpha$ fluctuations:} This has $0<\eta_\alpha < \eta_T$, 
so that $\eta_K$ and $\varepsilon_K$ are both positive.
From equation~(\ref{gamma}), the dimensional growth rate may be expressed as:
\beq
\gamma \;\;=\;\; \sigma\left\{\,
\left[\frac{\tau_\alpha}{\tau_*} \,-\, 1\right]
\left(\frac{k}{k_\alpha}\right)^2 \,-\,
\frac{\eta_\alpha}{\vert \eta_K \vert}
\left(\frac{k}{k_\alpha}\right)^4\right\}\,,
\label{gam-weak}
\eeq
\noindent
where the characteristic frequency $\sigma$ is defined earlier in
equation~(\ref{sigma}). We consider following two cases:
\begin{itemize}
\item[(a)]
$\underline{\mbox{Case}\;\tau_\alpha < \tau_*\,}$: 
In this case the growth rate $\gamma$ is negative at all wavenumbers as may be
seen from the solid curve in the left panel of Fig.~(\ref{Fig3}).
\item[(b)]
$\underline{\mbox{Case}\;\tau_\alpha > \tau_*\,}$: Here the growth rate is positive
for a range of wavenumbers and it is nonmonotonic function of $k$. Starting from
zero, it first increases with $k$, attains a maximum positive value,
\beq
\gamma_{\rm max} \,=\, \frac{\sigma \vert \eta_K \vert}{4\eta_{\alpha}}
\left[\frac{\tau_\alpha}{\tau_*} \,-\, 1\right]^2\;,\qquad
\mbox{at}\qquad \vert k \vert \,=\, k_{\rm max} \,=\,
k_{\alpha} \left[\frac{\vert \eta_K \vert}{2\eta_{\alpha}}
\left(\frac{\tau_\alpha}{\tau_*} \,-\, 1\right)\right]^{1/2}\,,
\eeq
\noindent
and then it decreases monotonically
for larger $k$, turning negative at sufficiently high wavenumber;
see bold curve in the right panel of Fig.~(\ref{Fig3}).
We note that \emph{the growth rate becomes negative at high enough
wavenumbers, thus exhibiting a high wavenumber cutoff, which would
enable a bonafide large--scale dynamo with suppression of magnetic power at
smaller scales.}
\end{itemize}

\begin{figure}
\begin{center}
\includegraphics[width=\columnwidth]{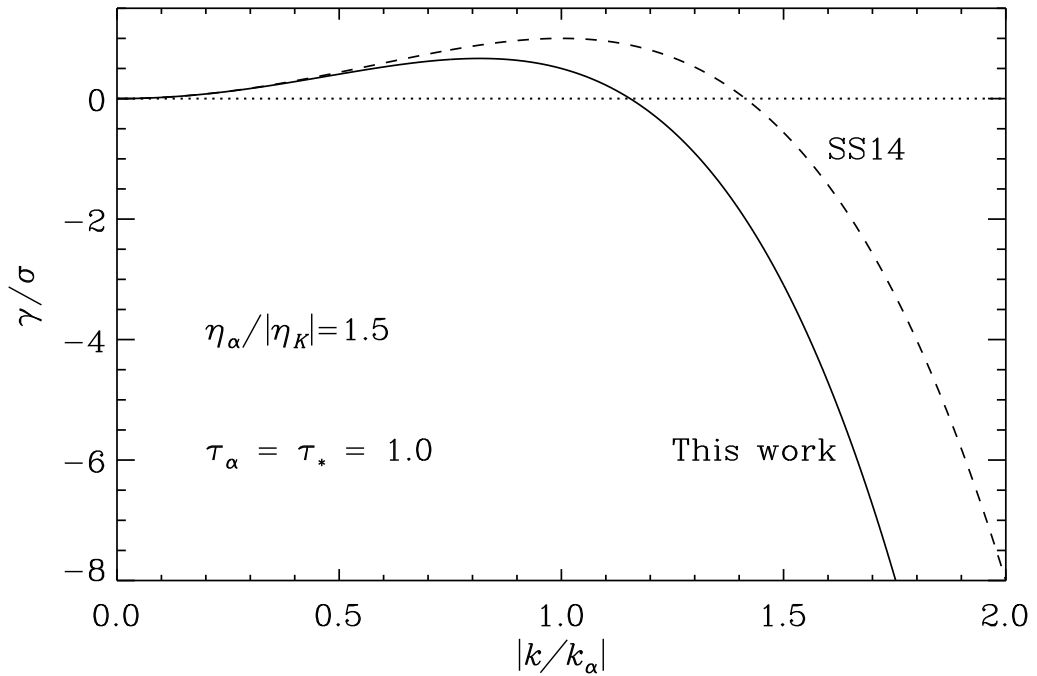}
\end{center}
\caption[]{Growth rate $\gamma/\sigma$ plotted as a function of
$\vert k/k_{\alpha}\vert$ for \emph{strong $\alpha$ fluctuations};
shown by bold curve. Dashed curve correspond to the results of
SS14.}
\label{Fig4}
\end{figure}

\noindent
{\bf Strong $\alpha$ fluctuations:} This has $0 < \eta_T < \eta_\alpha\,$, 
so that $\eta_K$ and $\varepsilon_K$ are both negative.
From equation~(\ref{gamma}), the dimensional growth rate may be expressed as:
\beq
\gamma \;\;=\;\; \sigma\left\{\,
\left[\frac{\tau_\alpha}{\tau_*} \,+\, 1\right]
\left(\frac{k}{k_\alpha}\right)^2 \,-\,
\frac{\eta_\alpha}{\vert \eta_K \vert}
\left(\frac{k}{k_\alpha}\right)^4\right\}\,.
\label{gam-strong}
\eeq
\noindent
As shown by solid curve in Fig.~(\ref{Fig4}), the growth rate $\gamma$ starts
from zero at $|k|=0$, increases with $k$ to reach a maximum positive value,
\beq
\gamma_{\rm max} \,=\, \frac{\sigma \vert \eta_K \vert}{4\eta_{\alpha}}
\left[\frac{\tau_\alpha}{\tau_*} \,+\, 1\right]^2\;,\qquad
\mbox{at}\qquad \vert k \vert \,=\, k_{\rm max} \,=\,
k_{\alpha} \left[\frac{\vert \eta_K \vert}{2\eta_{\alpha}}
\left(\frac{\tau_\alpha}{\tau_*} \,+\, 1\right)\right]^{1/2}\,,
\eeq
\noindent
beyond which it begins to decrease monotonically, and becomes negative for sufficiently
large wavenumbers.

In both Figs.~(\ref{Fig3}) and (\ref{Fig4}) we also show the corresponding
results of \cite{SS14} to illustrate some qualitative
differences.\footnote{We note, however, that the
expressions for the growth rates in SS14 were simpler and did not
explicitly depend on the factor $\eta_{\alpha}/|\eta_K|$.}
We notice good agreement at low wavenumbers whereas at large wavenumbers the
theory of \cite{SS14} overpredicts the growth rates. They had already pointed
out that such overestimation of growth/decay rates at large wavenumbers would be
expected as they dropped the diffusion term from the evolution equation for the
fluctuating magnetic field. In the present analysis where we retain this term,
we find that it affects the growth rate $\gamma$ in such
a way that it now always exhibits a
large wavenumber cutoff beyond which $\gamma$ becomes negative.
At large wavenumbers,
the new predictions for the growth rates are even qualitatively different
from the results of SS14 model and have only been possible due to
a nontrivial extension of the previous work.
\emph{Particularly interesting is the possibility of dynamo
action in case of
weak $\alpha$ fluctuations due to finite Moffatt drift,
where a window of small to intermediate wavenumbers
allows dynamo growth; see right panel of Fig.~(\ref{Fig3}).}

\section{Conclusions}

We have developed a theory of large--scale dynamo action where the mean magnetic
field grows solely due to an $\alpha$ parameter that is varying stochastically
in space and time with zero mean. Using the first--order--smoothing--approximation
or the quasi--linear approach,
we derived a closed integro--differential equation governing the evolution of the
large--scale magnetic field, which is non-perturbative in the $\alpha$--correlation
time ($\tau_{\alpha}$). This is the main result of this paper where we have
generalized the Kraichnan--Moffatt (KM) model \citep{Kra76,Mof78}, to include
effects of nonzero $\alpha$--correlation time, in a spirit similar to that of
\cite{SS14}. We, however, note that \cite{SS14} included shear in their analysis
while ignoring the diffusion term from fluctuating field equation, whereas here
we ignore shear but include the diffusion term, which is necessary for making
comparisons with results from future numerical experiments.
We show that statistically anisotropic $\alpha$ fluctuations give rise to a
drift velocity, called Moffatt drift, which contributes a new term in the mean EMF.
We first applied our model to the exactly solvable case of white--noise
($\tau_{\alpha}=0$) $\alpha$ fluctuations, in which case 
the mean EMF is identical to the KM model and the evolution
of the mean magnetic field depends on two constants, namely, the Kraichnan diffusivity
($\eta_K$) and the Moffatt drift ($\bfV_{\!\!M}$). We confirm earlier findings
\citep{Kra76,Mof78,SS14} that when $\tau_{\alpha}=0$,
(a) the necessary condition for dynamo action is that the fluctuations must be
strong, and (b) the Moffatt drift contributes only to the phase and
does not determine the growth or decay of the large--scale magnetic field.

In order to explore memory effects of fluctuating $\alpha$ on dynamo action,
we considered nonzero $\tau_{\alpha}$. Assuming that the $\tau_{\alpha}$ be
small and the large--scale magnetic field is slowly varying, we reduce the general
integro--differential equation to a partial differential equation
and state sufficient conditions for its
validity. Here each component of the mean magnetic field evolves independently
of the other components.
We provided an explicit expression for the growth rate
of the mean magnetic field and studied its behaviour
as a function of wavenumber ($k$), for different
choices of parameters involved.
Some salient results may be stated as follows:
\begin{enumerate}
 \item In the absence of the Moffatt drift, the necessary condition for dynamo
 action is that the $\alpha$ fluctuations must be strong. This is, in a sense,
 qualitatively similar to the white--noise or the original Kraichnan model, except
 that, here, we find the growth rate to be a nonmonotonic function of $k$, thus
 exhibiting a wavenumber cutoff beyond which it becomes progressively negative.
 \item For nonzero $\tau_{\alpha}$, the Moffatt drift contributes positively
 to the dynamo growth and it can always facilitate large--scale dynamo action
 if sufficiently large, even in case of weak $\alpha$ fluctuations.
 \item In the most general case when both, the Kraichnan diffusivity and the Moffatt
 drift, are nonzero, and $\tau_{\alpha}$ is finite, we find the possibility of
 dynamo growth in both regimes (weak and strong) of $\alpha$ fluctuations.
 We also determine the growth rate and corresponding wavenumber of the fastest
 growing mode.
 \item We find that there always exists a wavenumber cutoff at some large $k$
 beyond which the growth rate turns negative, irrespective of weak or strong
 $\alpha$ fluctuations. This makes it a special dynamo as the magnetic power
 at smallest length scales would be suppressed, thus enabling a bonafide
 large--scale dynamo.
\end{enumerate}

Thus a minimal extension of KM model to include effects of finite memory results
in a large--scale dynamo, driven by the Moffatt drift which arises in presence
of statistically anisotropic $\alpha$ fluctuations. Such a possibility was
first discussed in \cite{SS14}. It is particularly intriguing to find that
even weak $\alpha$ fluctuations could lead to the growth of mean magnetic
field due to finite Moffatt drift, where the maximum
growth occurs at intermediate length scales
($\sim$ few $k_{\alpha}$). Due to $k^2$ contribution to the growth rate, the
Moffatt drift driven dynamo appears to be of negatively diffusive type,
with coefficient of (negative) turbulent diffusion being
$V_{\!M3}^2 \tau_{\alpha}$.
However, it is different from the usual picture of negative diffusion
where the maximum growth occurs at smallest length scales.
Therefore, while the usual negative diffusion cannot continue
indefinitely in the mean--field framework for reasons stated
earlier, the Moffatt drift driven dynamo does not have such
limitations, resulting in a bonafide large--scale dynamo action.
This analysis leading to new contributions to the mean EMF
is expected to find applications in context to astrophysical dynamos, such as,
the disk dynamos, the solar dynamo etc.
Numerical as well as analytical explorations of
this new class of large--scale dynamos, by also considering
fluctuations in all components of tensorial transport coefficients
$\alpha_{ij}$ and $\eta_{ij}$, will be the
focus of a future investigation.

\section*{Acknowledgements}
I thank S. Sridhar for suggesting this problem and useful discussions.
Hospitality provided by RRI, Bangalore, where this work began during my visit
in November, 2014, is gratefully acknowledged. I also thank Luke Chamandy
for a critical reading of an earlier version of this manuscript
and for providing many useful suggestions.
I thank all the anonymous referees for many useful suggestions and
criticisms which have improved the quality of the present manuscript.

\appendix
\section{Derivation of Equations~(\ref{Q2}) and (\ref{C})}

To simplify the integral in equation~(\ref{Qdef}), let us first expand
$(\bfk-\bfk')^2 (\bfk-\bfk')$ in the integrand and write $\bfQ(\bfk)$ as,
\begin{eqnarray}
\bfQ(\bfk) &=& -\,\mathrm{i} \eta_T \Biggl\{
k^2\bfk \underbrace{\int \frac{\mathrm{d}^3k'}{(2\pi)^3}
\widetilde{{\cal A}}(\bfk')}_\text{I}
\,-\, k^2 \underbrace{\int \frac{\mathrm{d}^3k'}{(2\pi)^3} \bfk'
\widetilde{{\cal A}}(\bfk')}_\text{II}
\,-\, 2\bfk \underbrace{\int \frac{\mathrm{d}^3k'}{(2\pi)^3}
(\bfk\,\cendot\bfk') \widetilde{{\cal A}}(\bfk')}_\text{III}\nonumber\\[2ex]
&& + \bfk \underbrace{\int \frac{\mathrm{d}^3k'}{(2\pi)^3} k'^2
\widetilde{{\cal A}}(\bfk')}_\text{IV}
\,+\, 2 \underbrace{\int \frac{\mathrm{d}^3k'}{(2\pi)^3}
(\bfk\,\cendot\bfk') \bfk' \widetilde{{\cal A}}(\bfk')}_\text{V}
\,-\, \underbrace{\int \frac{\mathrm{d}^3k'}{(2\pi)^3} k'^2 \bfk'
\widetilde{{\cal A}}(\bfk')}_\text{VI}
\Biggr\}
\label{Q3}
\end{eqnarray}
\noindent
Writing the spatial correlation function as,
\beq
{\cal A}(\bfxi) = \int \frac{\mathrm{d}^3k'}{(2\pi)^3}\,
\exp{(\mathrm{i}\,\bfk'\cendot\bfxi)}\; \widetilde{{\cal A}}(\bfk')\;,
\eeq
\noindent
we can express the integrals I--VI in the equation~(\ref{Q3}) as following:
\begin{eqnarray}
&&\mbox{I}={\cal A}({\bf 0})\;;\qquad
\mbox{II}=-\mathrm{i}\Big[\bnabla {\cal A}(\bfxi)\Big]_{\bfxi={\bf 0}}\;;\qquad
\mbox{III}=-\mathrm{i}\bfk\,\cendot\Big[\bnabla {\cal A}(\bfxi)\Big]_{\bfxi={\bf 0}}
\nonumber \\[2ex]
\mbox{IV}&=&-\Big[\nabla^2 {\cal A}(\bfxi)\Big]_{\bfxi={\bf 0}}\;;\quad
\mbox{V}=-\Big[(\bfk\,\cendot\bnabla) \bnabla {\cal A}(\bfxi)\Big]_{\bfxi={\bf 0}}
\;;\quad
\mbox{VI}=\mathrm{i}\Big[\nabla^2\{\bnabla {\cal A}(\bfxi)\}\Big]_{\bfxi={\bf 0}}
\label{terms}
\end{eqnarray}
\noindent
Using (\ref{terms}) in (\ref{Q3}), we obtain the equation~(\ref{Q2}),
with definitions of $C_1$ (scalar), $\overleftrightarrow{\bfC_2}$ (dyad)
and $\bfC_3$ (vector) as given in the equation~(\ref{C}).
Also, recall that ${\cal A}({\bf 0})=\eta_{\alpha}$
and $-\Big[\bnabla {\cal A}(\bfxi)\Big]_{\bfxi={\bf 0}}=\bfV_{\!\!M}$.
Whereas our model does not specify the form of ${\cal A(\bfxi)}$ which is
by construct a large--scale quantity, but otherwise an arbitrary function,
we restrict our present analysis to the limit
$\ell_{\cal A} > \ell_{\overline{B}}$, where $\ell_{\cal A}$ and
$\ell_{\overline{B}}$ are typical scales of variation associated with
${\cal A}$ and large--scale magnetic field, respectively.
In this case the terms IV--VI involving $C$'s in the
above expressions can be safely ignored in comparison to the
rest of the terms in equation~(\ref{Q3}).

\section{Dimensionless growth rate function}

The expression for the growth rate as given in equation~(\ref{gamma}) may be
written in a dimensionless form using the parameters first defined in SS14:
\beq
\Gamma \;=\; \gamma \tau_\alpha\,,\qquad \beta = \eta_\alpha k^2\tau_\alpha\,,
\qquad \varepsilon_K \;=\; \eta_Kk^2\tau_\alpha\,,\qquad
\varepsilon_M \;=\; kV_{\!M3}\tau_\alpha\,.
\label{dimless}
\eeq
\noindent
There is just one constraint involving $\beta$ and $\varepsilon_K$,
coming from $\beta + \varepsilon_K = \eta_Tk^2\tau_\alpha > 0\,$.
Thus the parameter ranges are given by,
\beq
0 \;\leq\; \beta \;<\; \infty\,, \qquad \beta \;+\; 
\varepsilon_K \;>\; 0\,,
\qquad \vert\varepsilon_K\vert \;\ll\; 1\,,\qquad
\vert\varepsilon_M\vert \;\ll\; 1\,,
\label{parrange}
\eeq
\noindent
where the latter two conditions come from equation~(\ref{cond}).
Multiplying equation~(\ref{gamma}) by $\tau_\alpha > 0\,$, we obtain 
the dimensionless growth rates,
\beq
\Gamma \;=\; -\left(\varepsilon_K \,+\, \beta^2 \right) \;+\; \varepsilon_M^2
\label{Gamma}
\eeq
\noindent
as a function of the 3 dimensionless parameters
$\left(\beta\,, \varepsilon_K\,, \varepsilon_M\right)$.
Of these, the two parameters $\left(\varepsilon_K\,, \varepsilon_M\right)$
can be taken to be independently specified, taking positive 
and negative values, so long as their magnitudes are small.
But $\beta\geq 0\,$ is subject to the constraint $\beta + \varepsilon_K > 0\,$.
Therefore we can rewrite the conditions of (\ref{parrange}) as:
\begin{eqnarray}
&&\vert\varepsilon_K\vert \;\ll\; 1\,,\qquad\qquad
\vert\varepsilon_M\vert \;\ll\; 1\,,
\nonumber\\[1ex]
&& \mbox{For $\;\varepsilon_K \leq 0\,$, have
$\;\vert\varepsilon_K\vert \;<\; \beta \;<\; \infty\,$;
$\qquad$ For $\;\varepsilon_K > 0\,$, have
$\;0 \;\leq\; \beta \;<\; \infty\,$.}
\label{parrange2}
\end{eqnarray}

\begin{figure}
\begin{center}
\includegraphics[width=\columnwidth]{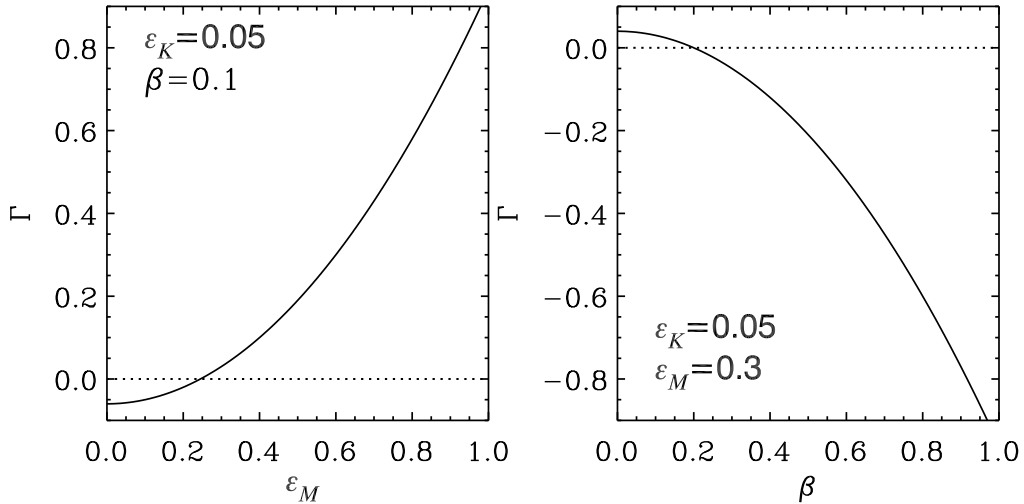}
\end{center}
\caption[]{Growth rate function $\Gamma$ plotted as a function of,
(i) $\varepsilon_M$, for $\beta=0.1$ (left panel), and (ii) $\beta$, for
$\vert\varepsilon_M\vert = 0.3$ (right panel); $\varepsilon_K=0.05$
in both the cases.}
\label{Fig2}
\end{figure}

\noindent
The dynamo condition is determined by a surface in three--dimensional parameter
space (spanned by $\beta\,, \varepsilon_K\,, \varepsilon_M$) at which
$\Gamma=0$, separating the dynamo region (with $\Gamma > 0$) from the
non--dynamo region (with $\Gamma < 0$).
In Fig.~(\ref{Fig2}) we plot the growth rate function $\Gamma$ as a
function of single parameter, keeping the other two parameters fixed;
left and right panels show $\Gamma$ as functions of $\varepsilon_M$
and $\beta$, respectively, with positive $\varepsilon_K\;(=0.05)$,
which corresponds to weak $\alpha$ fluctuations.
At fixed $\varepsilon_K$ and $\beta$, $\Gamma$ increases quadratically
with $\varepsilon_M$, whereas it decreases quadratically with
$\beta$ at fixed $\varepsilon_K$ and $\varepsilon_M$.
We note that the Moffatt drift, together with finite correlation time
of $\alpha$--fluctuations (parameterized by $\varepsilon_M$),
contributes positively to the dynamo growth and can always facilitate
large--scale dynamo action if sufficiently strong, even in case of
weak $\alpha$ fluctuations.

\section{Possible physical meanings of the parameters}

If the spatial correlation function ${\cal A}(\bfr)$ varies over scales
of order $\ell_{\cal A}$, then the Moffatt drift speed
$V_{\!M}\sim \eta_{\alpha}/\ell_{\cal A}$, giving from
equation~(\ref{taustar}),
$\tau_{*}\sim |f-1|\ell_{\cal A}^{\,2}/\eta_{\alpha}$, with factor
$f\equiv \eta_T/\eta_{\alpha}$. Note that $\tau_{*}$ is a
parameter which can be uniquely determined once the form of ${\cal A}(\bfr)$
is specified, whereas $\tau_{\alpha}$ is defined independently
by temporal correlation funtion ${\cal D}(t)$ using
equation~(\ref{corr-time}).
Two interesting limits can be sought: (i) when $\eta_T\ll \eta_\alpha$,
$\tau_{*}\sim \ell_{\cal A}^{\,2}/\eta_{\alpha}$, and (ii) when
$\eta_T\gg \eta_\alpha$,
$\tau_{*}\sim f\ell_{\cal A}^{\,2}/\eta_{\alpha}$, i.e., for weak
$\alpha$ fluctuations, it is larger by factor $f$ (with $f\gg 1$).
The wavenumber $k_\alpha$ as defined in equation~(\ref{kalpha})
signifies the inverse diffusion length due to $\alpha$--diffusivity
$\eta_\alpha$. Similarly the modified turbulent diffusion, or
Kraichnan diffusivity $\eta_K$ has an associated resistive scale
with corresponding wavenumber, say, $k_K=1/\sqrt{\eta_K \tau_\alpha}$.
The dimensionless parameters $\beta$ and $\varepsilon_K$
as defined in equation~(\ref{dimless})
characterise the wavenumbers of the modal mean--field solutions
in units of $k_\alpha$ and $k_K$, respectively, whereas $\varepsilon_M$
normalizes it by the distance traversed by Moffatt drift speed in time
$\tau_\alpha$.

\bibliographystyle{jfm}

\label{lastpage}

\end{document}